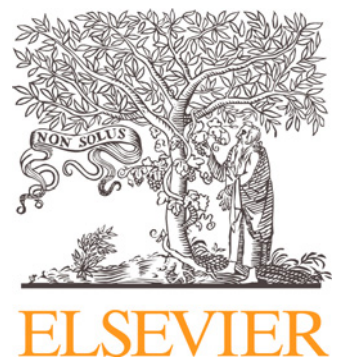

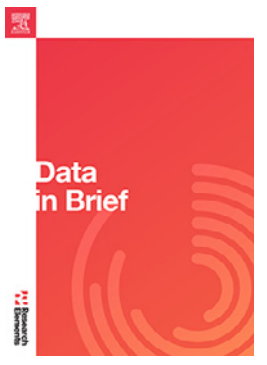

Data Article

# Dataset of total emissivity for $CO_2$, $H_2O$, and $H_2O$-$CO_2$ mixtures; over a temperature range of 300-2900 K and a pressure-pathlength range of 0.01-50 atm.m

Osama A. Marzouk

*College of Engineering, University of Buraimi, Al Buraimi, Postal Code 512, Oman*

A R T I C L E  I N F O

A B S T R A C T

This article describes a dataset of total (spectrally-integrated rather than wavelength-dependent) emissivity values pertaining to gaseous media containing carbon dioxide ($CO_2$) and/or water vapor ($H_2O$) with the sum of their partial pressures being near the atmospheric level. These conditions may be particularly relevant to flue gases resulting from oxy-fuel combustion. The emissivities here are computed using the EM2C implementations of the statistical narrow band (SNB), and they are made conveniently available as 10 separate plain text files having a unified layout. Each data file in the dataset corresponds to a specific chemical composition (from pure $CO_2$ to pure $H_2O$), with eight intermediate $H_2O$:$CO_2$ molar ratios being 1:20, 1:8, 1:4, 1:2, 1:1, 2:1, 5:1, and 20:1. In addition, pure $CO_2$ corresponds to the extreme lower-bound for the $H_2O$:$CO_2$ molar (as 0:1), and pure $H_2O$ corresponds to the extreme upper-bound for the $H_2O$:$CO_2$ molar ratio (1:0 or infinity). For each chemical composition, the total emissivity values are provided for different pressure-pathlengths and different absolute gas temperatures as two independent variables. The pressure-pathlength range spans about three orders of magnitude, from 0.01 atm.m to 50 atm.m, with 90 nonuniformly-distributed pressure-pathlength values. The absolute gas temperature spans a wide range from 300 K (room temperature) to 2900 K (high-temperature flames) with a uniform step of 25 K separating 105 temperature val-

*E-mail address:* osama.m@uob.edu.om

ues. For each data file, there are 90 × 105 (9450) emissivity values; and the entire dataset contains 94500 emissivity values.

Specifications Table

| | |
|---|---|
| Subject | Engineering & Materials science |
| Specific subject area | Thermal radiation property of a gaseous medium. |
| Type of data | Table.<br>Raw. |
| Data collection | The data were generated numerically based on a computer code that applies the EM2C implementation of the statistical narrow band (SNB) approach for estimating the total emissivity of a mixture of gases with optional soot. EM2C is a French acronym for "laboratoire d'Energetique Moléculaire et Macroscopique, Combustion", and its English translation is (Molecular and Macroscopic Molecular Energetics laboratory) [1]. The computer code is written in the FORTRAN programming language. We accessed this computer code among other codes that accompany a popular textbook about thermal radiation [2]. We compared results from this EM2C-SNB approach with other approaches to estimate the total emissivity, and the comparison provides confidence in the reported EM2C-SNB results [3]. |
| Data source location | The location in which the data were generated is Morgantown, West Virginia, USA. However, they are personal data (not institutional data). |
| Data accessibility | Repository name: Mendeley Data<br>Data identification number: 10.17632/x5wjzk6sjs.1 [4]<br>Direct URL to data: https://data.mendeley.com/datasets/x5wjzk6sjs/1<br>Instructions for accessing these data: none |
| Related research article | Osama A. Marzouk (2025). Technical review of radiative-property modeling approaches for gray and nongray radiation, and a recommended optimized WSGGM for CO2/H2O-enriched gases. Results in Engineering. 25:103923. 10.1016/j.rineng.2025.103923 [5]. |

## 1. Value of The Data

This dataset was originally intended as a training dataset for optimizing a weighted-sum-of-gray-gases model (WSGGM) that we proposed for modeling thermal radiation in oxy-fuel combustion or other industrial environments enriched with carbon dioxide ($CO_2$) and/or water vapor ($H_2O$) [6–8]. However, the dataset can be utilized in a variety of ways as described below:

- The dataset can be used as standalone lookup tables for the total emissivity as a discrete function of the local chemical composition, pressure-pathlength, and temperature; which can be used in computational fluid dynamics (CFD) simulations at elevated temperatures [9–11].
- The dataset can be used to replicate our published WSGGM model, and this gives other researchers confidence in their implemented algorithm and computational procedure, which they can subsequently extend.
- The dataset can be used as a training set for optimizing new non-conventional reduced-order models for the total emissivity, that can be used conveniently in lieu of a large set of discrete values.
- The dataset can be used as a benchmarking reference for assessing other approaches for estimating the total emissivity.
- The dataset can be used to gain insights into how the emissivity of the $CO_2$-$H_2O$ changes in a three-dimensional space formed by the chemical composition, pressure-pathlength, and gas temperature.

- The dataset can be used in deciding whether or not the thermal radiation is negligible (its submodel can be switched off) at the given problem condition, such as at a threshold low gas temperature.

Although other datasets related to the radiative emissivity of $CO_2$ and $H_2O$ exist, such as the HITRAN (HIgh-resolution TRANsmission molecular absorption database) database, and the HITEMP (HIgh-TEMPerature molecular database), these contain high-resolution spectroscopic line-by-line (LBL) information that needs to be processed through an intensive computational process before obtaining a single value of total emissivity for a gaseous mixture of $CO_2$ and $H_2O$ [12–16]. On the other hand, our dataset does not require any computations. The user can access directly the final emissivity results. There is also a nonlinear regression simple model to estimate the total emissivity for a gaseous mixture containing $CO_2$ and $H_2O$. However, it is applicable only to a temperature range from 1000 K to 2000 K, which is outside temperatures encountered in some combustion systems with elevated temperatures [17]. On the other hand, our dataset covers a wider range from 300 K to 2900 K.

## 2. Background

The weighted-sum-of-gray-gases model (WSGGM) is a mathematical model that approximates a radiatively-active gaseous medium by a number of hypothetical gray (spectrally-independent) gases, each of them has its own linear absorption coefficient (that depends on the pressure-pathlength) and temperature-dependent weights for its contribution to the total emissivity estimated for the real gaseous medium. Therefore, although a single real gas (or gas-mixture) is approximated by a number of fictitious gases, the level of complexities actually decreases because the real gas demands a complicated description of its spectral emissivity, while the emissivity of each of the multiple gray gases are much easier to describe. The values of the WSGGM parameters are optimized by minimizing the deviation between the predicted total emissivities and those provided as a training set over the range of pressure-pathlengths and gas temperatures of interest. For each chemical composition, a different set of parameters should be optimized. In our published research article linked with this data article, our proposed WSGGM has 240 parameters, which cover 10 chemical compositions of the radiatively-active gaseous species $CO_2$ and $H_2O$, with a varying molar ratio ($R = H_2O{:}H_2O$) that ranges from 0 (pure $CO_2$) to infinity (pure $H_2O$). In our published research article, the dataset we use in developing our WSGGM is not provided. By providing it here, the value of that published research article is increased through better replicability and enabled access to underlying raw data.

## 3. Data Description

The dataset of concern for the current data article has a simple file structure. It is composed of 10 plain-text data files, with no folders or subfolders. Furthermore, the 10 data files have a common text layout. Each data file has a size of 326 kB (kilobytes), and it can be viewed using a basic text editor/viewer. The name format of each data file is “R=xxxxxx_EM2C-SNB_totalEmissivities_90 × 105.dat”, where the placeholder “xxxxxx” designates the molar ratio R ($H_2O{:}CO_2$) to which this data file corresponds. For example, the data file “R=01.000_EM2C-SNB_totalEmissivities_90 × 105.dat” corresponds to the molar ratio of $R = 1.0$ (equal mole fractions of 50% for either $H_2O$ or $CO_2$). The numbers “90 × 105” in the data file name refers to the number of total emissivities in that data file, which is 9450, corresponding to 90 values of the pressure-pathlength (PL in atm.m) and 105 values of the absolute gas temperature (Tg, in K).

Table 1 summarizes the 10 data files, and the specific chemical composition of $H_2O$ and/or $CO_2$ to which the total emissivities reported in that file correspond. The total pressure in all cases is 1 atm; therefore, the mole fraction can also be viewed as a partial pressure expressed in standard atmospheres [18,19].

**Table 1**
Chemical compositions for the 10 data files in the dataset.

| Arbitrary file index | File name | R | Mole fraction of $H_2O$ | Mole fraction of $CO_2$ |
|---|---|---|---|---|
| 1 | R=00.000_EM2C-SNB_totalEmissivities_90 × 105.dat | 0 | 0% | 100% |
| 2 | R=00.050_EM2C-SNB_totalEmissivities_90 × 105.dat | 0.05 | 4.762% | 95.238% |
| 3 | R=00.125_EM2C-SNB_totalEmissivities_90 × 105.dat | 0.125 | 11.11% | 88.89% |
| 4 | R=00.250_EM2C-SNB_totalEmissivities_90 × 105.dat | 0.25 | 20% | 80% |
| 5 | R=00.500_EM2C-SNB_totalEmissivities_90 × 105.dat | 0.5 | 33.33% | 66.67% |
| 6 | R=01.000_EM2C-SNB_totalEmissivities_90 × 105.dat | 1 | 50% | 50% |
| 7 | R=02.000_EM2C-SNB_totalEmissivities_90 × 105.dat | 2 | 66.67% | 33.33% |
| 8 | R=05.000_EM2C-SNB_totalEmissivities_90 × 105.dat | 5 | 83.33% | 16.67% |
| 9 | R=20.000_EM2C-SNB_totalEmissivities_90 × 105.dat | 20 | 95.238% | 4.762% |
| 10 | R=Infinity_EM2C-SNB_totalEmissivities_90 × 105.dat | Infinity | 100% | 0% |

```
 number of PL[atm-m] values    number of Tg[K] values
    90                           105
 PLmin          PLmax        deltaPL_multiplicative
   0.01000      50.0000    1.10043
 Tgmin          Tgmax        deltaTg_additive
    300.00      2900.00    25.0000
 Ptot[atm]    XH2O[moleFrac]    XCO2[moleFrac]
    1.0000    0.500000000000    0.500000000000

    iPL     PL[atm-m]
     1       0.010000000
     2       0.011004276
     3       0.012109408
     4       0.013325527
     5       0.014663777
     6       0.016136424
     7       0.017756966
     8       0.019540255
     9       0.021502635
    10       0.023662092
```

**Fig. 1.** A view of the top part of one of the data files (R=01.000_EM2C-SNB_totalEmissivities_90 × 105.dat).

Given the common layout of the plain text within the data files, we explain this layout through one definite example we select, which is the data file “R=01.000_EM2C-SNB_totalEmissivities_90 × 105.dat”.

Fig. 1 illustrates the beginning of the example data file “R=01.000_EM2C-SNB_totalEmissivities_90 × 105.dat”. The content starts with useful metadata about the main data, with easily-understood field names.

In the first two lines of the data file content, the number of pressure-pathlength (PL) values is stated as 90, and the number of gas temperature (Tg) values is also stated as 105. In addition, the unit used for the pressure-pathlength (PL) is clarified as [atm.m], and the unit used for the gas temperature (Tg) is also clarified as [K].

In the 3rd and 4th lines, the range of the pressure-pathlength (PL) is specified by the minimum value (PLmin) and the maximum value (PLmax). In this dataset, PLmin = 0.01 atm.m, and PLmax = 50 atm.m. We discretize this very wide range (about three orders of magnitudes) as a geometric (multiplicative) series, rather than an arithmetic (additive) series for proper representation of the lower edge of this extensive range [20,21]. The multiplicative factor for this geometric series is specified as (deltaPL_multiplicative), with the value of 1.10043. This value is computed as

$$\Delta PL = \sqrt[89]{\frac{PL_{\max}}{PL_{\min}}} = 1.100427565 \tag{1}$$

where 89 is the number of nonuniform pressure-pathlength steps.

The 5th and 6th lines specify the maximum and minimum values of the gas temperature in the datafile (and the entire dataset); as Tgmin = 300 K and Tgmax = 2900 K, respectively. Unlike the pressure-pathlength, the range of the gas temperature is sufficiently bounded such that it can be reasonably covered with a uniform step, which is deltaTg_Additive = 25 K [22,23]. This is computed as

$$\Delta Tg = \frac{Tg_{\max} - Tg_{\min}}{104} = 25\text{K} \tag{2}$$

where 104 is the number of uniform temperature steps.

In the 7th and 8th lines, the total pressure, Ptot (in atm), is provided; and it is 1 atm in the entire dataset. It should be noted that this (Ptot) is the total pressure of the entire gaseous medium, which can be larger than the summed partial pressures of carbon dioxide ($CO_2$) and water vapor ($H_2O$), allowing additional radiatively-transparent gases to be present, such the hydrogen ($H_2$), oxygen ($O_2$), nitrogen ($N_2$), and argon (Ar) [24–26]. In the same two lines, the mole fraction of water vapor (XH2O) and mole fraction of carbon dioxide (XCO2) are specified as fractional numbers, not as percentages. In the example data file discussed here, these mole fractions are equally 0.5. The sum of the partial pressures of water vapor and carbon dioxide gives the “pressure” value in the pressure-pathlength variable. This is the partial pressure of the gases that are selective radiators, and this radiation-related pressure is computed as

$$P = P_{tot}\,(X_{H2O} + X_{CO2}) \tag{3}$$

where ($X_{H2O}$) is the mole fraction of $H_2O$, and ($X_{CO2}$) is the mole fraction of $CO_2$. In the current dataset, ($X_{H2O} + X_{CO2}$)=1.0; thus the gas medium is treated to be totally composed of $H_2O$, $CO_2$, or a mixture of both; and this is relevant to oxy-fuel combustion settings that facilitate carbon capture for mitigating $CO_2$ emissions [27,28].

The 9th line is an empty separator, and the lines 10-100 list the nonuniformly spaced PL values in the dataset, along with an integer index (iPL) for these 90 PL values (from 0.01 atm.m to 50 atm.m).

The content of the example data file “R=01.000_EM2C-SNB_totalEmissivities_90 × 105.dat” is continued in Fig. 2, which starts at line 95, near the end of the PL list. Then, line 101 is an empty separator. The total emissivity values start from line 102, and continue till the last line of data (line number 9552). Instead of displaying the total emissivity values as a large two-dimensional array to show the dependence on the pressure-pathlength (PL) and gas temperature (Tg), it is flattened and displayed as a sequence of one-dimensional column vectors. Each vector of total emissivity corresponds to one value of PL, but all the 105 values of Tg. In the subsequent vector, the next higher PL value is fixed, and all the 105 Tg values are encountered.

For each total emissivity, a row record is displayed with three entries. The PL value corresponding to that total emissivity appears in the first entry within the same row; the Tg value corresponding to that total emissivity appears in the second entry within the same row; and finally, the total emissivity appears in the third entry. Although this style of recording can be made compacter, we prefer this elaborate style, making each line an independent record that

```
85      30.985799098
86      34.097627457
87      37.521969161
88      41.290209164
89      45.436884337
90      50.000000000

PL[atm-m]   Tg[K]    total-emissivity[dimLESS]
 0.01000    300.00   0.114786276
 0.01000    325.00   0.108336920
 0.01000    350.00   0.103204567
 0.01000    375.00   0.099042698
 0.01000    400.00   0.095569625
 0.01000    425.00   0.092559337
 0.01000    450.00   0.089831173
 0.01000    475.00   0.087239390
 0.01000    500.00   0.084662903
 0.01000    525.00   0.083928484
 0.01000    550.00   0.083185825
```

**Fig. 2.** A view of an intermediate part of one of the data files (R=01.000_EM2C-SNB_totalEmissivities_90 × 105.dat).

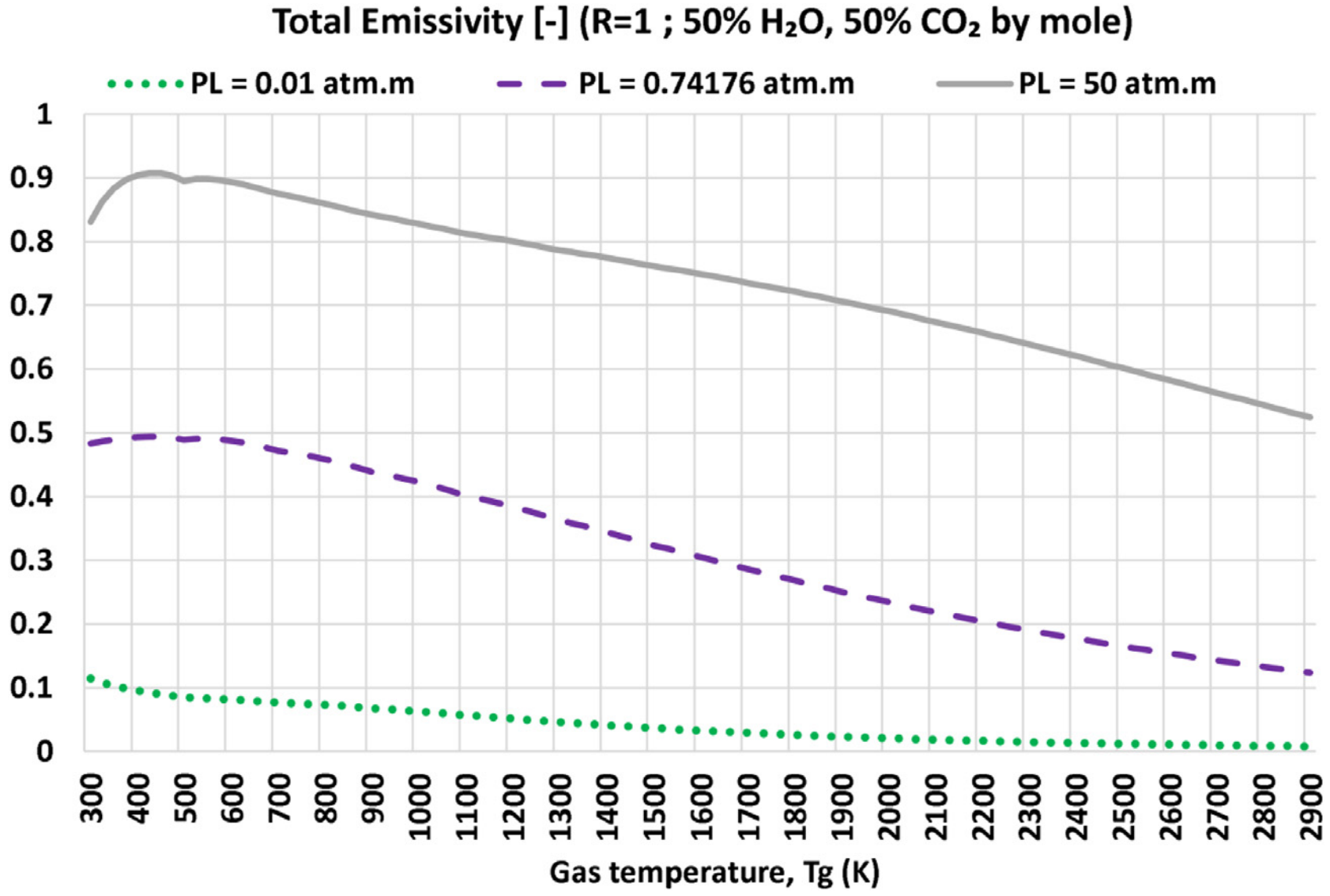

**Fig. 3.** Sampled one-dimensional profiles of the total emissivity of R ($H_2O$:$CO_2$ molar ratio) = 50%:50% as a function of gas temperature, obtained by processing the data file (R=01.000_EM2C-SNB_totalEmissivities_90 × 105.dat).

can be easily imported using spreadsheet software or a computer code, as well as interpreted by a human reader. In the particular example displayed here in the figure, the total emissivity of a 50%:50% (by mole or pressure) mixture of $H_2O$:$CO_2$ at 0.01 atm.m pressure-pathlength and 300 K is 0.114786276. The total emissivity values are reported with a fixed high precision (8 decimal places).

In Fig. 3, we provide an example of post-processing the dataset to demonstrate one of its useful uses through revealing some patterns of potential interest. In this example (corresponds to the same selected example data file “R=01.000_EM2C-SNB_totalEmissivities_90 × 105.dat”), the data is sampled at three different pressure-pathlength (PL) values, which are the minimum value (0.01 atm.m), the maximum value (50 atm.m), and an intermediate value of 0.74176 atm.m (close to the geometric mean of the minimum and the maximum values, which is 0.70711 atm.m). The variation of the total emissivity with the gas temperature is efficiently visualized in the figure, which shows different patterns that depend on the pressure-pathlength (PL) as well as on the gas temperature (Tg). Such nonlinear behavior is difficult (if even possible) to predict from analytical formulations. However, the provided dataset allows deeper investigations into the nonlinear influence of PL and Tg. For example, at the small PL value, the emissivity is not only small regardless of the temperature, but also decreases monotonically (but nonlinearly) with the temperature. For the larger PL values, the emissivity first increases with Tg, reaching a maximum, and then decreases.

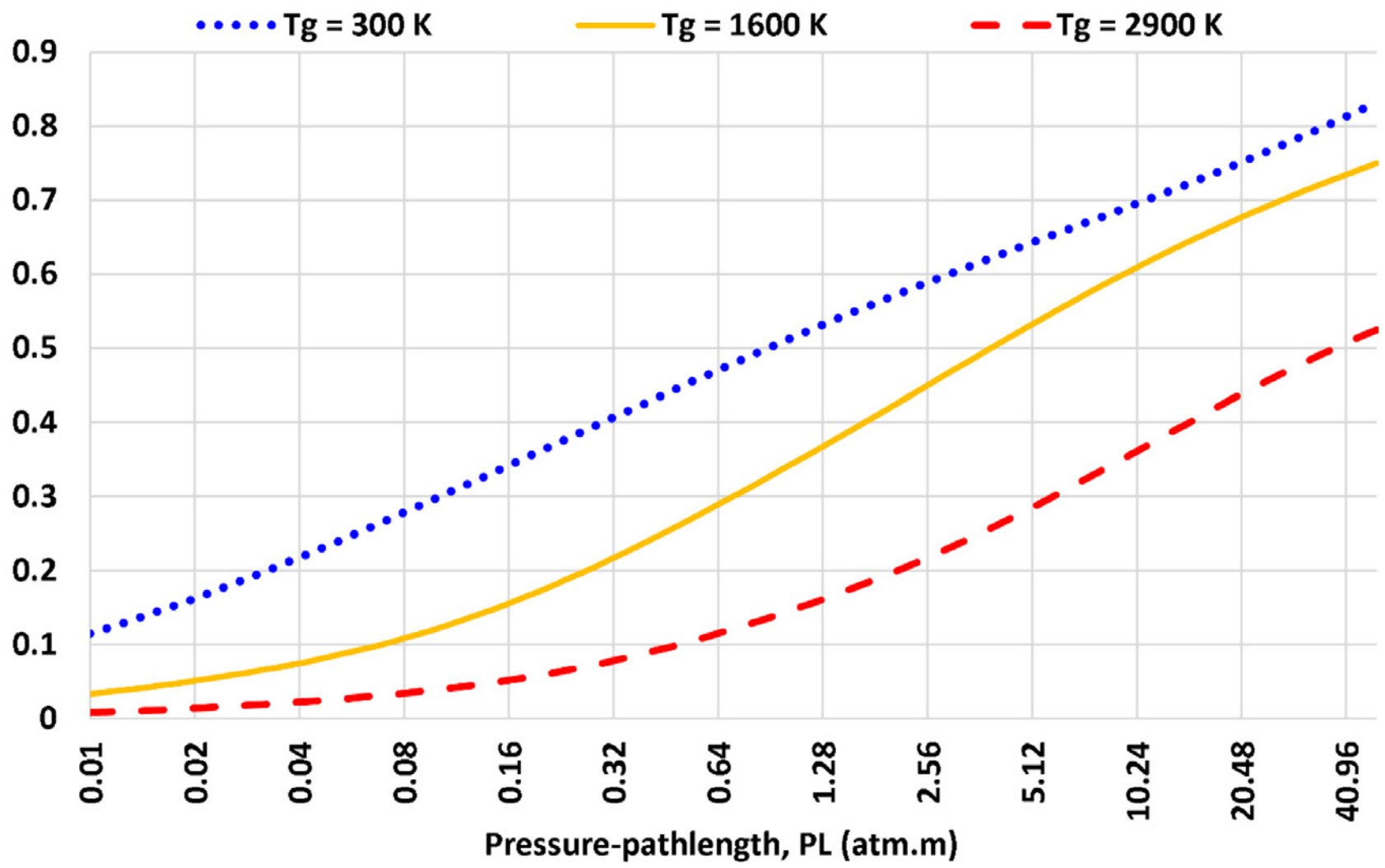

**Fig. 4.** Sampled one-dimensional profiles of the total emissivity of R ($H_2O$:$CO_2$ molar ratio) = 50%:50% as a function of the pressure-pathlength, obtained by processing the data file (R=01.000_EM2C-SNB_totalEmissivities_90 × 105.dat).

In Fig. 4, we provide additional extracted one-dimensional profiles of the total emissivity for $H_2O$:$CO_2$ = 1.0, but versus the pressure-pathlength (PL), at three different selected gas temperatures (Tg). These temperatures are the minimum value of 300 K, the maximum value of 2900 K, and the middle value of 1600 K. Overall, the total emissivity always increases as the temperature increases, but the rate and profile of this increase depends on the PL value.

If a linear absorption coefficient ($K_L$) with the unit of (1/m) is desired rather than the emissivity; for example, to solve the radiative transfer equation (RTE) in computational fluid dynamics (CFD) solvers; then this can be deduced from the total emissivity ($\in_{tot}$) in the dataset as [29–31]

$$K_L = -\frac{1}{L_m}\ln(1 - \in_{tot}) \tag{4}$$

where ($L_m$) is a mean characteristic pathlength suitable for the problem.

## 4. Experimental Design, Materials and Methods

The total emissivity dataset is not measured, but rather computed based on the FORTRAN computer code “SNB.F” that accompanies the reputable textbook in thermal radiation “Radiative Heat Transfer”[2]. The FORTRAN programming language has been used commonly in scientific and engineering applications [32]. It is a compiled language, which means that the computer code needs to be compiled using a FORTRAN compiler before it becomes ready for use. This allows the code to run fast [33]. We preferred not to include it with the public dataset because while we own the dataset, we do not own that computer code.

The computer code "SNB.F" has 422 lines, and it has several comment lines within it (comment lines start with the letter "C"), which help understand its algorithm. The interested reader is encouraged to consult the textbook for more details.

## Limitations

The dataset corresponds to the condition of an atmospheric total pressure (the total pressure is 1 atm). However, it should be noted that the adoption of an atmospheric pressure to construct such a gaseous emissivity dataset for flue gases is considered reasonable because an atmospheric pressure level is a suitable reference value for several energy and industrial applications, including combustion [34,35].

In addition, the dataset assumes a pure single-phase gaseous medium. The presence of soot is not considered in the dataset. Soot (or carbon black) is an undesirable byproduct of incomplete combustion, taking the form of black powder that tends to stick to solid surfaces [36]. However, it should be noted that there are several applications where the environment is a pure gaseous medium containing high concentrations of carbon dioxide and/or water vapor without soot, even when combustion is involved [37–39].

## Ethics Statement

The current work does not involve human subjects, animal experiments, or any data collected from social media platforms.

## Credit Author Statement

**Osama A. Marzouk:** Conceptualization, Methodology, Software, Validation, Formal analysis, Investigation, Writing - Original Draft, Writing - Review & Editing, Visualization.

## Data Availability

Emissivity_CO2-H2O_94500-values_EM2C (Original data) (Mendeley Data).

## Acknowledgements

This research did not receive any specific grant from funding agencies in the public, commercial, or not-for-profit sectors.

## Declaration of Competing Interest

The author declares that they have no known competing financial interests or personal relationships that could have appeared to influence the work reported in this paper.